\documentclass[letterpaper,10pt,conference]{ieeeconf}
\IEEEoverridecommandlockouts                              
\usepackage{graphicx}
\usepackage{amsmath,amssymb,latexsym,mathtools,mathrsfs}
\usepackage{algorithm}
\usepackage{algpseudocode}
\usepackage{multirow}
\usepackage{xcolor}

\makeatletter
\let\NAT@parse\undefined
\makeatother

\usepackage{hyperref}
\hypersetup{%
  colorlinks=true,%
  linkcolor={red!50!black},
  citecolor={blue!65!black},
  urlcolor={blue!80!black},
  bookmarksnumbered=true,%
  bookmarksopen=true}

\usepackage{cite}

\newcommand{\mydot}{\bullet}
\newcommand{\htwo}{\ensuremath{\text{H}_2} }
\newcommand{\heh}{\ensuremath{\text{HeH}^+} }
\newcommand{\ba}{\ensuremath{\mathbf{a}} }
\newcommand{\bc}{\ensuremath{\mathbf{c}} }

\newcommand{\bbf}{\ensuremath{\mathbf{f}} }

\newcommand{\balpha}{\ensuremath{\boldsymbol{\alpha}} }
\newcommand{\bbeta}{\ensuremath{\boldsymbol{\beta}} }
\newcommand{\blambda}{\ensuremath{\boldsymbol{\lambda}} }
\newcommand{\btheta}{\ensuremath{\boldsymbol{\theta}} }
\newcommand{\bmu}{\ensuremath{\boldsymbol{\mu}} }

\title{\LARGE \bf
Second-Order Adjoint Method for Quantum Optimal Control
}

\author{Harish S. Bhat
\thanks{This research was sponsored by the Office of Naval Research (Grant Number W911NF-23-1-0153).  The views and conclusions contained in this document are those of the authors and should not be interpreted as representing the official policies, either expressed or implied, of the Army Research Office or the U.S. Government. The U.S. Government is authorized to reproduce and distribute reprints for Government purposes notwithstanding any copyright notation herein.  This research used the Delta advanced computing and data resource which is supported by the U.S. National Science Foundation (award OAC 2005572) and the State of Illinois. Delta is a joint effort of the University of Illinois Urbana-Champaign and its National Center for Supercomputing Applications. This work used the Delta system at NCSA through allocation MTH230003 from the Advanced Cyberinfrastructure Coordination Ecosystem: Services \& Support (ACCESS) program, which is supported by U.S. National Science Foundation grants 2138259, 2138286, 2138307, 2137603, and 2138296 \cite{Boerner2023}.}%
\thanks{H. S. Bhat is with the Department of Applied Mathematics, University of California, Merced, Merced, CA 95343 USA 
        {\tt\small hbhat@ucmerced.edu}}%
}

\begin{document}

\maketitle
\thispagestyle{empty}
\pagestyle{empty}

\begin{abstract}
We derive and implement a second-order adjoint method to compute exact gradients and Hessians for a prototypical quantum optimal control problem, that of solving for the minimal energy applied electric field that drives a molecule from a given initial state to a desired target state.  For small to moderately sized systems, we demonstrate a vectorized GPU implementation of a second-order adjoint method that computes both Hessians and gradients with wall times only marginally more than those required to compute gradients via commonly used first-order adjoint methods.  Pairing our second-order adjoint method with a trust region optimizer (a type of Newton method), we show that it outperforms a first-order method, requiring significantly fewer iterations and wall time to find optimal controls for four molecular systems.  Our derivation of the second-order adjoint method allows for arbitrary parameterizations of the controls.
\end{abstract}


\section{INTRODUCTION}
\label{sect:intro}
In quantum optimal control, unless our problem admits a solution that we can derive by hand, we compute solutions using numerical optimization.  Within the space of gradient-based optimization methods, there is a broad distinction between (i) first-order methods that require only first derivatives of the objective and (ii) second-order methods that require first and second derivatives.  Many if not most numerical studies in quantum optimal control use first-order methods, which we take to include quasi-Newton methods that use gradients to update an approximate (inverse) Hessian.  In this paper, we derive both first- and second-order adjoint methods for a prototypical quantum optimal control problem.  The second-order adjoint method gives us an algorithm to compute exact Hessians, which enables use of an exact trust region solver, a second-order Newton method, to compute optimal controls.  We show that the second-order method yields more rapid convergence than the first-order method, both in terms of iteration count and wall clock time. 

We consider the dynamics of electrons in molecules with nuclei held fixed as part of the Born-Oppenheimer approximation.  The resulting dynamics are governed by the time-dependent Schr\"odinger equation (TDSE) with a core Hamiltonian consisting of a sum of kinetic operators (for the electrons), electron-nuclear potentials, and electron-electron potentials.  To this core Hamiltonian we add time-dependent terms that model applied electric fields within the dipole approximation.  In general, for a many-electron system, one cannot solve the resulting TDSE for the full time-dependent wave function without making approximations.  Here we employ a method equivalent to full configuration interaction, explained in Section \ref{sect:problem}, through which one can derive from the TDSE a tractable system of ordinary differential equations (ODE).  This ODE system is generic and arises whenever one expresses the TDSE's operators and wave function in a finite basis.

For a given molecule, we seek the optimal electric field that one should apply to transfer the molecule from a given initial state to a desired target state.  We relax this problem slightly and use a cost that combines the squared norm of the control effort (i.e., the energy of the applied field) with the squared error between achieved and target states at the final time.  This allows for the possibility that the desired target is unreachable.  The aim in such problems is to reach a final state that is sufficiently close to the desired target.

Though the mathematical form of our quantum optimal control problem can be found throughout the literature \cite{Werschnik_2007,Borzi2017,boscainIntroductionPontryaginMaximum2021,d2007introduction}, relatively few prior studies use exact Hessians and Newton-type methods.  Among those that do, one strand incorporates exact Hessians into the GRAPE (gradient ascent pulse engineering) optimal control algorithm \cite{khaneja2005}, using either an auxiliary matrix exponential  \cite{GoodwinKuprov2015,GoodwinKuprov2016,goodwinAcceleratedNewtonRaphsonGRAPE2023} or exact diagonalization  \cite{Dalgaard2020} to compute the required second derivatives of the matrix exponential.  Another strand employs Krylov-Newton methods \cite[\S 6.3]{Borzi2017}, including a matrix-free, semismooth Newton approach in which the whole Hessian matrix need not be stored \cite{ciaramellaNewtonMethodsOptimal2015}. As in these works, we solve a quantum optimal control problem using an adjoint method that enforces the equations of motion (as equality constraints) via time-dependent Lagrange multipliers.  Alternatively, one can enforce these constraints via projection; \cite{Shao2024} combines such an approach with a Hessian-based Newton method.   

Our work adds to this literature in three ways.  First, throughout all prior work of which we are aware, the second-order adjoint method has not been used to derive algorithms to compute Hessians for quantum optimal control problems.  This is in contrast to optimal control, data assimilation, and inverse problems in other areas of science, for which second-order adjoint methods have proven useful, even for large-scale problems \cite{PetraSachs2021}.  Second, we implement our second-order adjoint method in JAX \cite{jax2018github}, taking advantage of GPU vectorization.  As we show, this leads to near-linear scaling of Hessian wall clock time as a function of the number of time steps, mitigating a common concern that Hessians take too long to compute.  Third, prior work has not fully explored different parameterizations of the open-loop control $\bbf(t)$.   We present a derivation of the second-order adjoint method that can be applied with arbitrary parameterizations of the control.  All code developed for this paper is available here: \url{https://github.com/hbhat4000/hessQOC}.

\section{PROBLEM FORMULATION}
\label{sect:problem}
For the electron dynamics problem described in Section \ref{sect:intro}, we apply the CASSCF method \cite{Roos1980} with a large enough active space such that it is equivalent to full configuration interaction (CI) \cite{szabo2012modern,mcweeny1989methods,BhatJMP2025}.  To describe full CI in brief, we begin with a finite basis set of atomic orbitals, each a spatial function of three variables.  From this basis set, one constructs one-electron molecular spin-orbitals (each a function of three spatial variables and spin), from which one builds Slater determinants.  Each Slater determinant is a function of the spatial and spin coordinates of all $N_{\text{e}}$ electrons in the molecular system.  Consider the set of all nontrivial Slater determinants.  A linear combination from this set is determined by a coefficient vector $\bc$.  The action of the time-independent core Hamiltonian on such a linear combination reduces to a Hermitian matrix $\mathcal{M}$ multiplying $\bc$.  We can use the eigendecomposition of $\mathcal{M}$ to form the full CI basis functions $\{\Psi^\text{CI}_{q} (\mathbf{X})\}_{q=1}^{N}$, each a linear combination of Slater determinants.  The full CI basis set is orthonormal \emph{and} diagonalizes the core Hamiltonian.

We expand the TDSE wave function in this basis, with time-dependent coefficient vector $\ba(t)$:
\begin{equation}
\label{eqn:TDCIexpan}
\Psi(\mathbf{X},t) = \sum_{q=1}^{N} a_q(t) \Psi^\text{CI}_{q}(\mathbf{X}).
\end{equation}
Using this in the TDSE, and working in atomic units where $\hbar = 1$, we obtain the ODE system
\begin{equation}
\label{eqn:matrixTDSE}
i \frac{d}{dt} \ba(t) = \biggl( H_0 + \sum_{k=1}^3 f^k(t) M_k \biggr) \ba(t).
\end{equation}
Here $H_0$ is the core Hamiltonian matrix, diagonal by the construction above.  In the $x$, $y$, and $z$ directions,  respectively, we have dipole moment matrices $\{M_1, M_2, M_3\}$ and applied electric field strengths $\bbf(t) = (f^1(t), f^2(t), f^3(t))$.  The $M_k$ matrices express the dipole moment operators in the full CI basis; in general, $M_k$ does not commute with $H_0$.  Because we have used all nontrivial Slater determinants, the only error incurred by replacing the TDSE with the finite-dimensional system (\ref{eqn:matrixTDSE}) is due to the finiteness of our atomic orbital basis set.  As the size of this basis set goes to infinity, we recover the exact solution of the TDSE with electronic Born-Oppenheimer Hamiltonian \cite{townsend2019post}.

\emph{The field strengths $\bbf(t)$ are the open-loop controls that we solve for in this work.}  We can now formulate our problem in continuous time: given $H_0$, $\{M_k\}_{k=1}^3$, a cost balance parameter $\rho > 0$, a final time $T > 0$, an initial state $\balpha$, and a target state $\bbeta$, solve for $\bbf : [0,T] \to \mathbb{R}^3$ that minimizes
\begin{equation}
\label{eqn:continuouscost}
C[f] = \frac{1}{2} \sum_{k=1}^3 \int_0^{T} f^k(t)^2 \, dt + \frac{\rho}{2} \| \ba(T) - \bbeta \|_2^2
\end{equation}
subject to the equation of motion (\ref{eqn:matrixTDSE}) and initial condition $\ba(0) = \balpha$.  Note that we have incorporated the target state into the cost; unlike a hard constraint of the form ``$\ba(T) = \bbeta$,'' we allow for the possibility that it may not be possible to exactly reach $\bbeta$ in time $T$ starting from $\balpha$.

To avoid gradient/Hessian inconsistencies that may arise in an optimize-then-discretize approach \cite[\S 6.3.3]{Borzi2017}, we take a discretize-then-optimize approach.  Using fixed time step $\Delta t > 0$ such that $J \Delta t = T$, we discretize (\ref{eqn:matrixTDSE}) via
\begin{subequations}
\label{eqn:eom}
\begin{align}
\label{eqn:Zdef}
Z_j &= -i\biggl(H_0 + \sum_{k=1}^3 f^k(j \Delta t; \btheta) M_k \biggr) \Delta t \\
\label{eqn:discTDSE}
\ba_{j+1} &= \exp(Z_j) \ba_{j}
\end{align}
\end{subequations}
Here $\ba_{j}$ approximates $\ba(j \Delta t)$ and $\exp$ denotes the matrix exponential.  We have introduced the \emph{model} $\bbf(j \Delta t; \btheta)$ with parameters $\btheta \in \mathbb{R}^{N_p}$, which we take to be our decision variables.  Aside from sufficient smoothness to compute required derivatives, we impose no other requirements on this model. 

Given $\{H_0, \{M_k\}_{k=1}^3, \rho, J, \Delta t, \balpha, \bbeta\}$, the discrete-time optimal control problem is to find $\btheta \in  \mathbb{R}^{N_p}$ that minimizes
\begin{equation}
\label{eqn:discretecost}
\mathscr{C}(\btheta) = \frac{1}{2} \sum_{k=1}^3 \sum_{j=0}^{J-1} f^k(j \Delta t; \btheta)^2 + \frac{\rho}{2} \| \ba_{J} - \bbeta \|_2^2
\end{equation}
subject to $\ba_0 = \balpha$ and (\ref{eqn:eom}) for $j=0, \ldots, J-1$.

In this paper, we focus on the two-electron molecules $\htwo$ and $\heh$ treated using both the STO-3G \cite[\S 3.6.2]{szabo2012modern} and 6-31G \cite{Hehre1972} basis sets.  For these molecules, full CI in STO-3G and 6-31G results in problem sizes of $N = 4$ and $N=16$, respectively. 
Also, for these molecules, the dipole moment matrices in the $x$ and $y$ directions vanish; the remaining control is  $f_3(t) \in \mathbb{R}$.

\section{SOLUTION METHODS}

\subsection{First-Order Adjoint Method}
The method begins with a Lagrangian that incorporates the discrete-time cost (\ref{eqn:discretecost}) and equations of motion (\ref{eqn:eom}):
\begin{multline}
\label{eqn:lag}
\mathscr{L}(A, \Lambda, \btheta) = \frac{1}{2} \sum_{k=1}^3 \sum_{j=0}^{J-1} f_j^k(\btheta)^2 + \frac{\rho}{2} \| \ba_{J} - \bbeta \|_2^2 \\
- \Re \sum_{j=0}^{J-1} \blambda_{j+1}^\dagger ( \ba_{j+1} - \exp(Z_j) \ba_j ),
\end{multline}
where $Z_j$ was defined in (\ref{eqn:Zdef}) and $f^k_j(\btheta) = f^k(j \Delta t; \btheta)$. Here $A = \{ \ba_j \}_{j=1}^J$ and $\Lambda = \{ \blambda_j \}_{j=1}^{J}$ denote the collections of states and costates, respectively.  In (\ref{eqn:lag}), we set $\ba_0 = \balpha$.   Our goal is to find a critical point of $\mathscr{L}$, i.e., to satisfy a necessary condition for the solution of the discrete-time optimal control problem.  Hence we compute gradients of (\ref{eqn:lag}) with respect to $A$, $\Lambda$ and $\btheta$.  As $\nabla_{\Lambda} \mathscr{L} = \mathbf{0}$ will reproduce (\ref{eqn:discTDSE}), we focus on the remaining two gradients.  Treating $\ba_k$ and $\ba_k^\ast$ as separate variables, we compute and set $\nabla_{\ba_k} \mathscr{L} = 0$, resulting in
\begin{subequations}
\label{eqn:lambdasys}
\begin{align}
\label{eqn:lambdafc}
\blambda_J &= \ba_J - \bbeta \\
\label{eqn:lambdaevolve}
\blambda_k^\dagger &= \blambda^\dagger_{k+1} \exp(Z_k).
\end{align}
\end{subequations}
This system determines $\blambda$: we start with the final condition (\ref{eqn:lambdafc}) and iterate (\ref{eqn:lambdaevolve}) backwards in time for $k = J-1, \ldots, 1$.  Next, we have for $k = 0, \ldots, J-1$:
\begin{multline}
\label{eqn:dLdf}
\frac{\partial \mathscr{L}}{\partial \theta_{\ell}} = \sum_{k=1}^3 \sum_{j=0}^{J-1} f^k_j(\btheta) \frac{\partial f^k_j}{\partial \theta_{\ell}} \\
- \Re \sum_{j=0}^{J-1} \blambda_{j+1}^\dagger \Biggl[ i \Delta t \sum_{k=1}^{3}  \frac{d}{dZ} \exp(Z) \biggr|_{Z = Z_j} \! \! \! \! \! \mydot M_k \frac{\partial f^k_j}{\partial \theta_{\ell}}  \Biggr] \ba_j .
\end{multline}
We assume that $\nabla_{\btheta} \bbf$ can  be computed efficiently either by hand or by automatic differentiation.  Prior work has shown how to compute the directional derivative $d (\exp Z)/dZ \mydot W$ for Hermitian or anti-Hermitian matrices $Z$ \cite{Lewis2001,GoodwinKuprov2015,Dalgaard2020}.  We have our own derivations, omitted due to space constraints, that result in efficient implementations.  For small to moderately sized $Z$, we compute $\exp(Z)$ by first computing the eigendecomposition $Z = V D V^\dagger$, which yields  $\exp(Z) = V \exp(D) V^\dagger$.  Once we have computed $V$ and $D$, we compute $d (\exp Z)/dZ \mydot W$ with low additional cost.

Starting with the right-hand side of $\mathscr{C}(\btheta)$ in (\ref{eqn:discretecost}), if we use (\ref{eqn:discTDSE}) to unwind $\ba_J$ all the way back to $\ba_0 = \balpha$, we obtain
\begin{equation}
\label{eqn:costunwind}
\mathscr{C}^\text{u}(\btheta) = \frac{1}{2} \sum_{j=0}^{J-1} \| \bbf_j(\btheta) \|_2^2 + \frac{\rho}{2} \Biggl\| \prod_{j=0}^{J-1} \exp(Z_j) \balpha - \bbeta \Biggr\|_2^2.
\end{equation}
The product is ordered so that as $j$ increases, successive terms \emph{left}-multiply prior terms.  As we have substituted all constraints into the cost, an equivalent formulation of our discrete-time optimal control problem is to minimize (\ref{eqn:costunwind}) over $\btheta$.  If we inspect (\ref{eqn:eom}) and (\ref{eqn:lambdasys}), we find that their respective solutions clearly depend on $\btheta$ through $Z_j$ defined in (\ref{eqn:Zdef}); we write this dependence as $A(\btheta)$ and $\Lambda(\btheta)$, respectively.  In the first-order adjoint method, given $\btheta$, we first solve (\ref{eqn:eom}) for $A(\btheta)$, and then solve (\ref{eqn:lambdasys}) for $\Lambda(\btheta)$.  Using these pieces to compute $\nabla_{\btheta} \mathscr{L}$ via (\ref{eqn:dLdf}), we will find
\begin{equation}
\label{eqn:gradagree}
\nabla_{\btheta} \mathscr{C}^\text{u}(\btheta) = \nabla_{\btheta} \mathscr{L}(A(\btheta),\Lambda(\btheta),\btheta).
\end{equation}
In words, the first-order adjoint method yields the same gradient as that obtained by differentiating the unconstrained cost (\ref{eqn:costunwind}).  Equipped with this gradient, we can iteratively update $\btheta$ with the goal of converging to a critical point.  We summarize the needed steps in Algorithm \ref{alg:firstorder}.

\begin{algorithm}[t]
\caption{First-order adjoint method to solve the discrete-time optimal control problem from Section \ref{sect:problem}}\label{alg:firstorder}
{\small \begin{algorithmic}[1]
\Require $\{H_0, \{M_k\}_{k=1}^3, \rho, J, \Delta t, \balpha, \bbeta\}$, $\btheta^{(0)}$, and $m=0$.
\State $\ba_0 \gets \balpha$
\For{$j = 0, \dots, J-1$}
\State Compute and store the decomposition $V_j D_j V_j^\dagger = Z_j$
\State $\ba_{j+1} \gets V_j \exp( D_j )V_j^\dagger \ba_j$
\EndFor
\State $\blambda_J \gets \ba_J - \bbeta$
\For{$k = J-1, \ldots, 1$}
\State $\blambda_k^\dagger \gets \blambda^\dagger_{k+1} V_k \exp(D_k) V_k^\dagger$
\EndFor
\State Use the stored eigendecomposition of $Z_k$ to compute and store $d (\exp Z)/dZ \mydot M$, evaluated at each $Z = Z_k$.
\State Using $\{\ba_j\}_{j=0}^{J-1}$, $\{\blambda_j\}_{j=1}^{J}$, the derivatives computed in line 10, and $\nabla_{\btheta} \bbf$, compute $\nabla_{\btheta} \mathscr{L}$ via (\ref{eqn:dLdf}).  Use $\nabla_{\btheta} \mathscr{L}$ in a gradient-based optimization method to compute the next iterate $\btheta^{(m+1)}$.
\State If $\|\btheta^{(m+1)} - \btheta^{(m)}\| < \delta$ or $\| \nabla_{\btheta} \mathscr{L}(\btheta^{(m+1)}) \| < \epsilon$, exit; else $m \gets m+1$ and return to line $1$.
\end{algorithmic}}
\end{algorithm}

\subsection{Second-Order Adjoint Method}
Let us start from the end: our goal in this subsection is to compute the Hessian of the unconstrained cost: $\nabla_{\btheta} \nabla_{\btheta} \mathscr{C}^\text{u}$.  From (\ref{eqn:gradagree}), we see that in order to compute the required Hessian from the Lagrangian $\mathscr{L}$, we must take a \emph{total} derivative of (\ref{eqn:dLdf}) with respect to $\btheta$, resulting in
\begin{multline}
\label{eqn:d2Ldf2}
\frac{d}{d {\theta}_{m}} \frac{\partial \mathscr{L}}{\partial \theta_{\ell}} = \sum_{k=1}^3 \sum_{j=0}^{J-1}
f^k_j(\btheta) \frac{\partial^2 f^k_j}{\partial \theta_{\ell} \partial \theta_{m}} + \frac{\partial f^k_j}{\partial \theta_{\ell}} \frac{\partial f^k_j}{\partial \theta_{m}} \\
+ \Re  \sum_{j=0}^{J-1} \Biggl\{ \frac{d \blambda_{j+1}^\dagger}{d \theta_m} \Biggl[ i \Delta t \sum_{k=1}^{3}  \frac{d}{dZ} \exp(Z) \biggr|_{Z = Z_j} \! \! \! \! \! \mydot M_k \frac{\partial f^k_j}{\partial \theta_{\ell}}  \Biggr] \ba_j \\
+ \blambda_{j+1}^\dagger \Biggl[  (\Delta t)^2 \! \! \sum_{k,n=1}^{3}  \frac{d}{dZ} \exp(Z) \biggr|_{Z = Z_j} \! \! \! \! \! \! \! \! \! \! \mydot (M_k \otimes M_n) \frac{\partial f^k_j}{\partial \theta_{\ell}} \frac{\partial f^n_j}{\partial \theta_{m}}  \Biggr] \ba_j \\
+ \blambda_{j+1}^\dagger \Biggl[ i \Delta t \sum_{k=1}^{3}  \frac{d}{dZ} \exp(Z) \biggr|_{Z = Z_j} \! \! \! \! \! \mydot M_k \frac{\partial^2 f^k_j}{\partial \theta_{\ell} \partial \theta_{m}}  \Biggr] \ba_j \\
+ \blambda_{j+1}^\dagger \Biggl[ i \Delta t \sum_{k=1}^{3}  \frac{d}{dZ} \exp(Z) \biggr|_{Z = Z_j} \! \! \! \! \! \mydot M_k \frac{\partial f^k_j}{\partial \theta_{\ell}}  \Biggr] \frac{d \ba_j}{d \theta_m} \Biggr\}.
\end{multline}
We see $\nabla_{\btheta} \bbf$ and  $\nabla_{\btheta} \nabla_{\btheta} \bbf$ throughout; we assume both  can  be computed efficiently either by hand or by automatic differentiation.  In the remaining lines, we see three new objects. One new object is the second derivative $d^2 (\exp Z)/dZ^2 \mydot (W_1 \otimes W_2)$. As before, we have derived expressions for this that can be implemented efficiently assuming we have precomputed $Z = V D V^\dagger$.  Another new object in (\ref{eqn:d2Ldf2}) is  $\bmu_{j,\ell} = d \blambda_j / d {\theta}_{\ell}$; to derive a scheme to compute this, we begin with the total derivative of the Lagrangian:
\begin{multline}
\label{eqn:gradlag}
\mathscr{L}'_{\ell} := \frac{d \mathscr{L}}{d {\theta}_{\ell}} = \sum_{k=1}^3 \sum_{j=0}^{J-1} f^k_j(\btheta) \frac{\partial f^k_j}{\partial \theta_{\ell}} + \rho \Re \biggl[ (\ba_{J} - \bbeta)^\dagger \frac{d \ba_{J}}{d {\theta}_{\ell}} \biggr] \\
- \Re \sum_{j=0}^{J-1} \biggl[ \bmu^\dagger_{j+1,\ell} (\ba_{j+1} - \exp(Z_j) \ba_j) \\
+ \blambda_{j+1}^\dagger \biggl( \frac{d \ba_{j+1}}{d f_{\ell}} -  \frac{d \exp(Z_j)}{d {\theta_{\ell}}}  \ba_j
- \exp(Z_j) \frac{d \ba_j}{d f_{\ell}} \biggr) \biggr]
\end{multline}
Note that $\mathscr{L}'$ defined by (\ref{eqn:gradlag}) is a function of $A = \{\ba_j\}_{j=1}^J$, $\Lambda = \{\blambda_j\}_{j=1}^{J}$ and $\nabla_{\btheta} \Lambda = \{\bmu_j = \nabla_{\btheta} \blambda_j \}_{j=1}^J$. As before, we seek a critical point of $\mathscr{L}'$.  Note also that $\nabla_{\bmu_k} \mathscr{L}' = \mathbf{0}$ and $\nabla_{\blambda_k} \mathscr{L}' = \mathbf{0}$ will reproduce both (\ref{eqn:discTDSE}) and the total derivative of (\ref{eqn:discTDSE}) with respect to $\btheta$.  Hence we focus on variations of $\mathscr{L}'$ with respect to $\ba$; with $I_{S}$ as the indicator variable for the set $S$, we have
\begin{multline}
\label{eqn:varLprime}
\delta \mathscr{L}'_{\ell} = \frac{d}{d \epsilon} \biggr|_{\epsilon=0}  \! \! \! \! \! \! \! \mathscr{L}'_{\ell} \bigl( \{ \ba_j + \epsilon \delta \ba_j \}_{j=1}^J \bigr) 
= \Re \sum_{j=0}^J \frac{ d \delta \ba_j}{d {\theta}_{\ell}} \\
\times \biggl[ \rho(\ba_J - \bbeta)^\dagger I_{\{j = J\}} - \blambda_j^\dagger I_{\{j \geq 1\}}  
+ \blambda_{j+1}^\dagger \exp(Z_j) I_{\{j \leq J-1\}} \biggr]  \\
+ \Re \sum_{j=0}^J \delta \ba_j \biggl[ \rho \frac{d \ba_J^\dagger}{d {\theta}_{\ell}} I_{\{j=J\}} - \bmu^\dagger_{j,\ell} I_{\{j \geq 1\}} \\
+ \bmu_{j+1,\ell}^\dagger \exp(Z_j) I_{\{j \leq J-1\}} + \blambda^\dagger_{j+1} \frac{d \exp(Z_j)}{d {\theta}_{\ell}}    \biggr] 
\end{multline}
The right-hand side of (\ref{eqn:varLprime}) contains two summations.  Requiring that the first summation vanishes for all variations $\nabla_{\btheta} \delta \ba_j$, we recover the first-order adjoint system (\ref{eqn:lambdasys}).  For the second summation to vanish for all variations $\delta \ba_j$, $\bmu$ must satisfy the second-order adjoint system with final condition $\bmu_{J,\ell} = \rho {d \ba_J}/d {\theta}_{\ell}$ and, for $j < J$,
\begin{multline}
\label{eqn:musys}
\bmu^\dagger_{j,\ell} = \bmu^\dagger_{j+1,\ell} \exp(Z_j) \\
- \blambda^\dagger_{j+1} 
 i \Delta t \sum_{k=1}^{3}  \frac{d}{dZ} \exp(Z) \biggr|_{Z = Z_j} \! \! \! \! \! \mydot M_k \frac{\partial f^k_j}{\partial \theta_{\ell}}.
\end{multline}
The last new object in both (\ref{eqn:musys}) and (\ref{eqn:d2Ldf2}) is $d \ba_j / d {\theta}_{\ell}$. With the total derivative of (\ref{eqn:discTDSE}) with respect to $\btheta$, we have
\begin{equation}
\label{eqn:gradatheta}
\frac{d \ba_{j+1}}{d \theta_{\ell}} = \exp(Z_j) \frac{d \ba_j}{d \theta_{\ell}} + i \Delta t \sum_{k=1}^{3}  \frac{d \exp(Z) }{dZ} \biggr|_{Z = Z_j} \! \! \! \! \! \! \! \! \! \! \mydot M_k \frac{\partial f^k_j}{\partial \theta_{\ell}} \ba_j.
\end{equation}
As $\ba_0 = \balpha$ does not depend on $\btheta$ at all, the gradient is zero: $d \ba_0 / d \theta_{\ell} = \mathbf{0}$.  With this initial condition, we iterate (\ref{eqn:gradatheta}) from $j=0$ to $j=J-1$ to compute all required gradients $\nabla_{\theta} \ba_j$.

In Algorithm \ref{alg:secondorder}, we collect all  results from this subsection and give an end-to-end procedure to iteratively solve for $\theta$ (and thereby solve for $\bbf$) using exact Hessians and gradients. 

\begin{algorithm}[t]
\caption{Second-order adjoint method to solve the discrete-time optimal control problem from Section \ref{sect:problem}}\label{alg:secondorder}
{\small \begin{algorithmic}[1]
\Require $\{H_0, \{M_k\}_{k=1}^3, \rho, J, \Delta t, \balpha, \bbeta\}$, $\btheta^{(0)}$, and $m=0$.
\State $\ba_0 \gets \balpha$
\For{$j = 0, \dots, J-1$}
\State Compute and store the decomposition $V_j D_j V_j^\dagger = Z_j$
\State $\ba_{j+1} \gets V_j \exp( D_j )V_j^\dagger \ba_j$
\EndFor
\State $\blambda_J \gets \ba_J - \bbeta$
\For{$j = J-1, \ldots, 1$}
\State $\blambda_j^\dagger \gets \blambda^\dagger_{j+1} V_j \exp(D_j) V_j^\dagger$
\EndFor
\For{$j = 0, \ldots, J-1$}
\State Use the eigendecomposition of $Z_j$ to compute and store all derivatives $d (\exp Z)/dZ \mydot M_k$ and $d^2 (\exp Z)/dZ^2 \mydot (M_k \otimes M_n)$, evaluated at $Z = Z_j$.
\EndFor
\For{$\ell = 1, \ldots, N_p$} \Comment{Parallelize/vectorize over $\ell$}
\State $d \ba_0 / d \theta_{\ell} \gets \mathbf{0}$
\For{$j = 0, \ldots, J-1$}
\State $d \ba_{j+1}/d \theta_{\ell} \gets V_j \exp(D_j) V_j^\dagger {d \ba_j}/{d \theta_{\ell}}$
\State $ - i \Delta t \sum_{k=1}^{3}  \frac{d}{dZ} \exp(Z)  \biggr|_{Z = Z_j} \! \! \! \! \! \mydot M_k (\partial f^k_j / \partial \theta_{\ell}) \ba_j$
\EndFor
\EndFor
\For{$\ell = 1, \ldots, N_p$} \Comment{Parallelize/vectorize over $\ell$}
\State $\bmu_{J,\ell} \gets \rho (d \ba_J / d f_{\ell})$
\For{$j = J-1, \ldots, 1$}
\State $\bmu^\dagger_{j,\ell} \gets \bmu^\dagger_{j+1,\ell} V_j \exp(D_j) V_j^\dagger$
\State $ -  i \Delta t \blambda^\dagger_{j+1} 
\sum_{k=1}^{3}  \frac{d}{dZ} \exp(Z) \biggr|_{Z = Z_j} \! \! \! \! \! \mydot M_k (\partial f^k_j/\partial \theta_{\ell})$ 
\EndFor
\EndFor
\State Using $\{\ba_j\}_{j=0}^{J-1}$, $\{\blambda_j\}_{j=1}^{J}$, $\{\nabla_{\btheta} \ba_j\}_{j=0}^{J-1}$, $\{\bmu_j\}_{j=1}^{J}$ and derivatives of the matrix exponential, compute $\nabla_{\btheta} \mathscr{L}$ via (\ref{eqn:dLdf}) and $\nabla_{\btheta} \nabla_{\btheta} \mathscr{L}$ via (\ref{eqn:d2Ldf2}); use the computed gradient and Hessian in a second-order optimization method to compute the next iterate $\btheta^{(m+1)}$.
\State If $\|\btheta^{(m+1)} - \btheta^{(m)}\| < \delta$ or $\| \nabla_{\btheta} \mathscr{L}(\btheta^{(m+1)}) \| < \epsilon$, exit; else $m \gets m+1$ and return to line 1.
\end{algorithmic}}
\end{algorithm}

\subsection{Operational Difference}
\label{sect:opdiff}
The first $9$ lines of both algorithms are identical.  The differences between the two algorithms are that Alg. \ref{alg:secondorder} requires (i) computation of second derivatives of the matrix exponential, (ii) two nested loops, from lines 13-19 and lines 20-26, and (iii) computation of the Hessian via (\ref{eqn:d2Ldf2}).  Regarding (i), let us assume that if one has already diagonalized each $Z_j$, then the cost of computing second derivatives of $\exp(Z)$ at $Z_j$ is a constant multiple of the cost of computing first derivatives of $\exp(Z)$ at $Z_j$.

As for (ii), in the two nested loops, the inner loops are very similar, each requiring a total of $4J$ matrix-vector multiplications and $J$ matrix-scalar multiplications.  The outer loops over the $\ell$ variable can be parallelized easily.  Here $\ell$ goes from $1$ to $N_p$, the length of the $\btheta$ vector, the total number of parameters in the model.  When we examine (iii), we see that computation of both the Hessian (\ref{eqn:d2Ldf2}) and the gradient (\ref{eqn:dLdf}) can also be parallelized over the $\ell$ variable.  In the above, we use \emph{parallelize} to include either shared-memory multiprocessing (e.g., via OpenMP) or vectorization.  Our implementations, designed to run on graphical processing units (GPUs), utilize vectorization via \texttt{vmap} constructs in JAX \cite{jax2018github}.  With effective parallelization or vectorization, we hypothesize that one pass of  either Alg. \ref{alg:firstorder} or \ref{alg:secondorder} will have a wall clock time that is nearly linear (if not linear) in $J$, the total number of time steps required to go from $t=0$ to $t=T$.  Below we will test this empirically.

\subsection{Models}
\label{sect:models}
Continuing the discussion of models $\bbf(j \Delta t; \btheta)$ from Section \ref{sect:problem}, we set
\begin{equation}
\label{eqn:maximal}
f^k(j \Delta t; \btheta) = \theta_{k,j}
\end{equation}
for a $3 \times J$ parameter matrix $\btheta$; we then recover a model in which we optimize directly over all individual values of the discrete-time controls $f^k(j \Delta t)$, equivalent to the piecewise constant model employed in prior work \cite{GoodwinKuprov2016,Dalgaard2020}. This model maximizes flexibility: any discrete-time control signal $\bbf(j \Delta t; \btheta)$ is achievable under this model.  Hence  $N_p = 3J$ is an upper bound on the number of model parameters. We explore the maximal model (\ref{eqn:maximal}) in several tests below.  

In future work, we will demonstrate the use of neural network models that take $t$ as input and produce as output $\bbf(t; \btheta)$.  Here  $\btheta$ is a concatenation of all neural network parameters (e.g., weights and biases).  By choosing the network architecture carefully, we can ensure $N_p = | \btheta | \ll 3J$, potentially making such models attractive when $J$ is large.

\begin{figure}[t]
\begin{center}\includegraphics[width=3in]{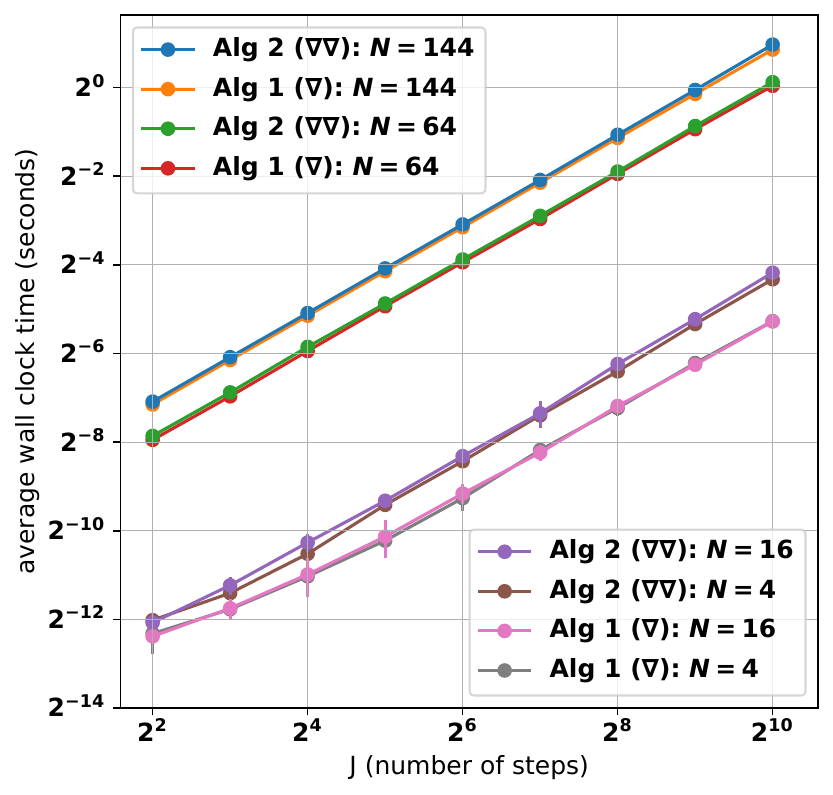}\end{center} \vspace{-0.5cm}
\caption{We give a log-log plot of wall clock time results for Algorithms \ref{alg:firstorder} and \ref{alg:secondorder}.  Here $N$ represents problem size, while $J$ is the number of time steps required to go from $t=0$ to $t=T$.  For each fixed choice of parameters $(N,J)$, we ran each algorithm $1000$ times; we plot the mean (circular dot) and error bar (plus/minus twice the standard deviation) of these runs.  \emph{The results show that both algorithms' scaling as a function of $J$ is close to linear---see Section \ref{sect:scaling} in the main text for details.}  The upshot: for the quantum optimal control problem we studied, Hessians and gradients can be computed at similar cost to gradients alone.}
\label{fig:timings}
\end{figure}

\begin{table*}[th]
\begin{center}
\begin{tabular}{lllrrrrrr}
 molecule   &  basis set    &   algorithm   &   \# iterations &   wall time &   final cost &   final $\| \text{grad} \|$ &   $\| \text{target viol} \|$ \\
\hline
 \multirow{ 2}{*}{$\htwo$} & \multirow{ 2}{*}{STO-3G} & Alg 1 &   5748.12 &          73.4884 &      \textbf{12.7697} &        0.882974    &         1.604$\times 10^{-5}$    \\
   & & Alg 2 &   \textbf{1421.67} &          \textbf{21.3366} &      12.7795 &        $\mathbf{6.2616 \times 10^{-5}}$  &         $\mathbf{1.57277 \times 10^{-5}}$  \\ \hline
  \multirow{ 2}{*}{$\htwo$}  & \multirow{ 2}{*}{6-31G}  & Alg 1 &   5313.17 &          92.6164 &      14.5274 &        1.28098     &         $\mathbf{2.41829\times 10^{-5}}$  \\
   & & Alg 2 &   \textbf{1335.56} &          \textbf{21.41}   &      \textbf{14.5221} &        $\mathbf{4.75845 \times 10^{-5}}$ &         2.52149$\times 10^{-5}$  \\ \hline
 \multirow{ 2}{*}{$\heh$} & \multirow{ 2}{*}{STO-3G} & Alg 1 &   7707.69 &          92.0181 &      21.0352 &        2.38341     &         3.23744$\times 10^{-5}$     \\
   & & Alg 2 &   \textbf{2729.95} &          \textbf{42.2176} &      \textbf{20.8101} &        $\mathbf{5.45808 \times 10^{-2}}$   &         $\mathbf{3.22242\times 10^{-5}}$     \\ \hline
  \multirow{ 2}{*}{$\heh$}   & \multirow{ 2}{*}{6-31G}  & Alg 1 &   6967.62 &         105.222  &      16.1364 &        3.25426     &         1.93098$\times 10^{-5}$     \\
   & & Alg 2 &   \textbf{1997.94} &          \textbf{31.8765} &      \textbf{15.8801} &        $\mathbf{8.06679 \times 10^{-3}}$  &         $\mathbf{1.77365\times 10^{-5}}$     \\
\end{tabular}\end{center}
\caption{Mean performance of each algorithm across $1000$ trials using the maximal model (\ref{eqn:maximal}).  The results show that Alg. \ref{alg:secondorder} outperforms Alg. \ref{alg:firstorder}, consistently requiring fewer iterations \emph{and} less wall clock time to achieve solutions of the same or better quality.  For details, see Section \ref{sect:ocr} in the main text.  For each molecular system, we boldface the best results.}
\label{tab:rawocr}
\end{table*}

\begin{table*}[th]
\begin{center}
\begin{tabular}{lllllll}
\hline
 molecule  & basis set   & \# iterations  & wall time     & final cost        & final $\| \text{grad} \|$ &   $\| \text{target viol} \|$  \\
\hline
$\htwo$    & STO-3G  & 5.08 (2.12, 9.88)  & 4.44 (1.72, 9.06)  & 1.00 (0.83, 1.17) & 5.00$\times 10^{6}$ (1.62$\times 10^{-1}$, 4.11$\times 10^{4}$) & 1.03 (0.83, 1.24)   \\
$\htwo$   & 6-31G   & 4.95 (1.66, 10.27) & 5.46 (1.58, 13.74) & 1.03 (0.68, 1.49) & 3.53$\times 10^{6}$ (2.84$\times 10^{-1}$, 1.20$\times 10^{5}$) & 1.05 (0.47, 1.98)   \\
$\heh$  & STO-3G  & 3.33 (1.67, 6.05)  & 2.58 (1.31, 4.81)  & 1.02 (0.81, 1.24) & 1.80$\times 10^{7}$ (2.87$\times 10^{-1}$, 1.72$\times 10^{7}$) & 1.03 (0.66, 1.52)   \\
$\heh$  & 6-31G   & 4.33 (1.56, 8.80)  & 4.15 (1.36, 9.14)  & 1.03 (0.80, 1.29) & 1.64$\times 10^{7}$ (3.83$\times 10^{-1}$, 5.85$\times 10^{6}$) & 1.12 (0.71, 1.79)   \\
\hline
\end{tabular}
\end{center}
\caption{Ratios indicating the relative performance of Alg. \ref{alg:firstorder} to Alg. \ref{alg:secondorder} using the maximal model (\ref{eqn:maximal}).  Note that ratios are computed first (in each of $1000$ trials), after which we compute means and $(0.05,0.95)$ quantiles, reported in parentheses.  For details, see Section \ref{sect:ocr} in the main text.  The results  show that using exact Hessians allows for more rapid convergence to solutions of substantially the same quality in terms of final cost and norm discrepancy between achieved and desired targets.}
\label{tab:ratios}
\end{table*}

\begin{figure*}[th]
{\centering \includegraphics[width=1.705in]{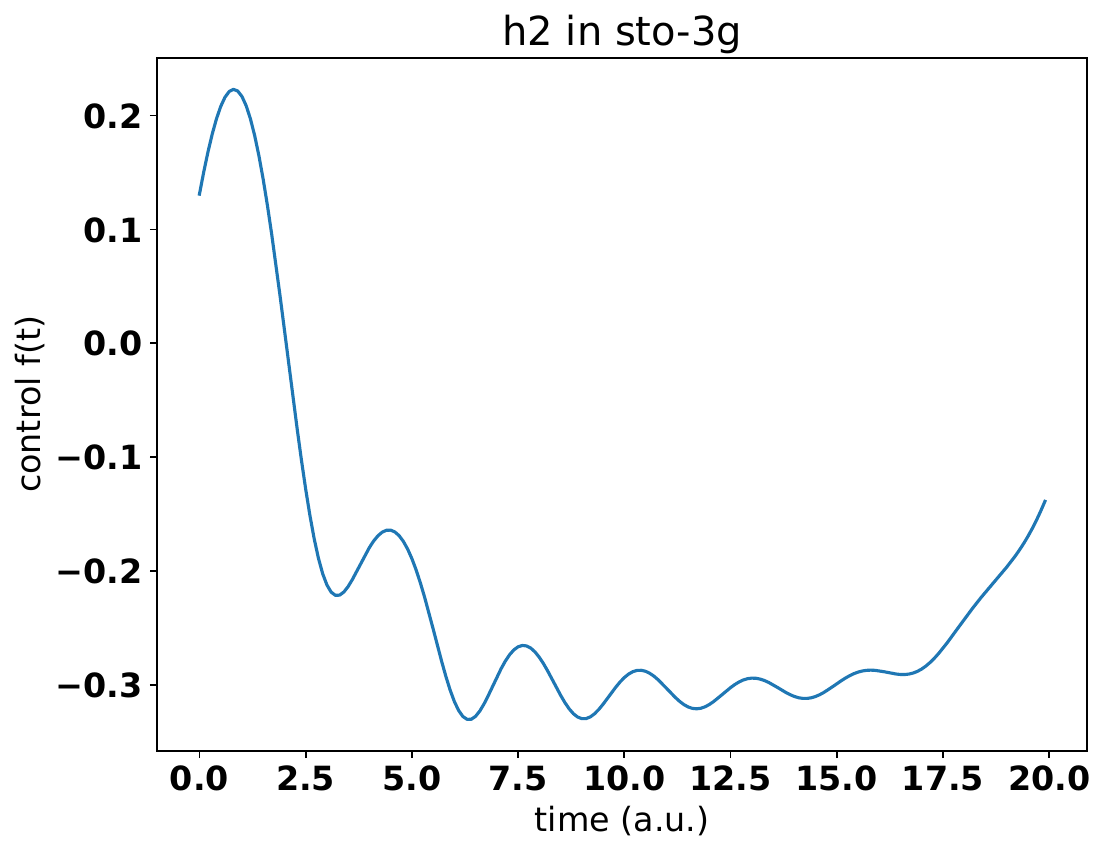} \hspace{-0.2cm} \includegraphics[width=1.795in]{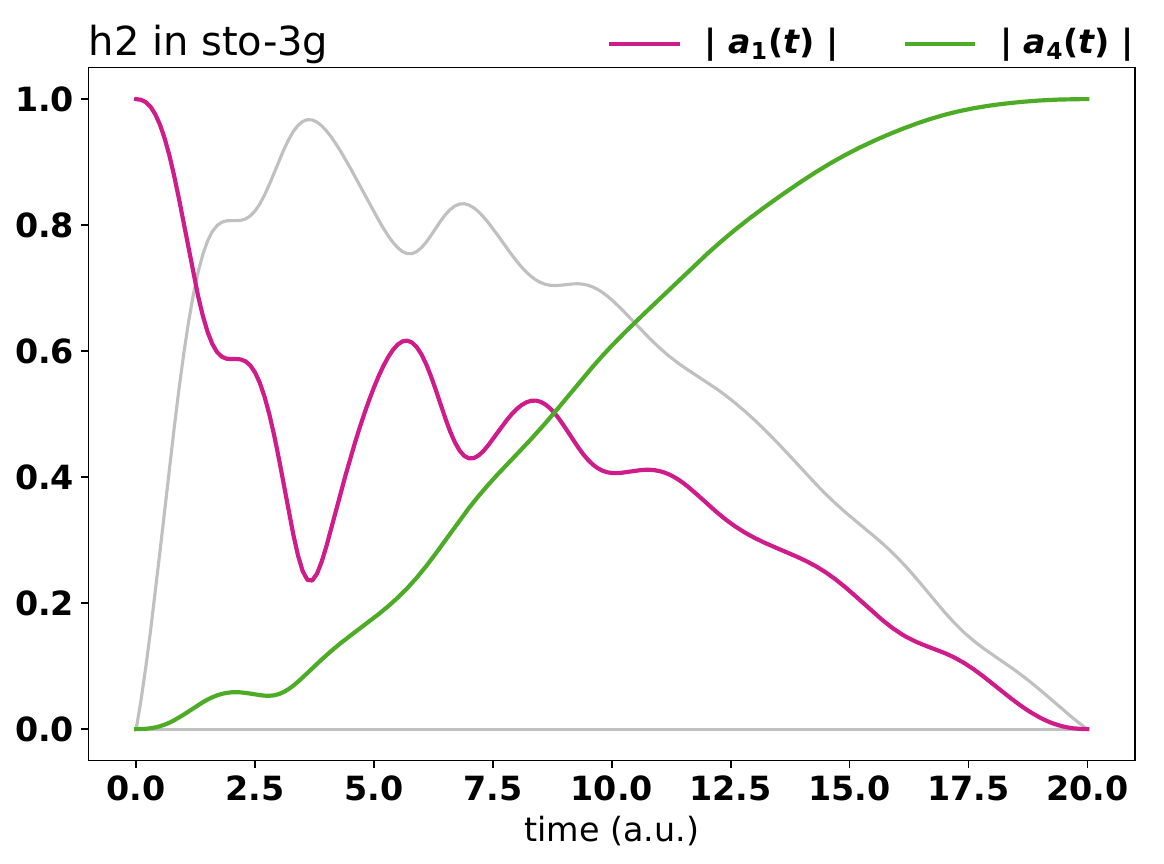} \hspace{-0.3cm} \includegraphics[width=1.705in]{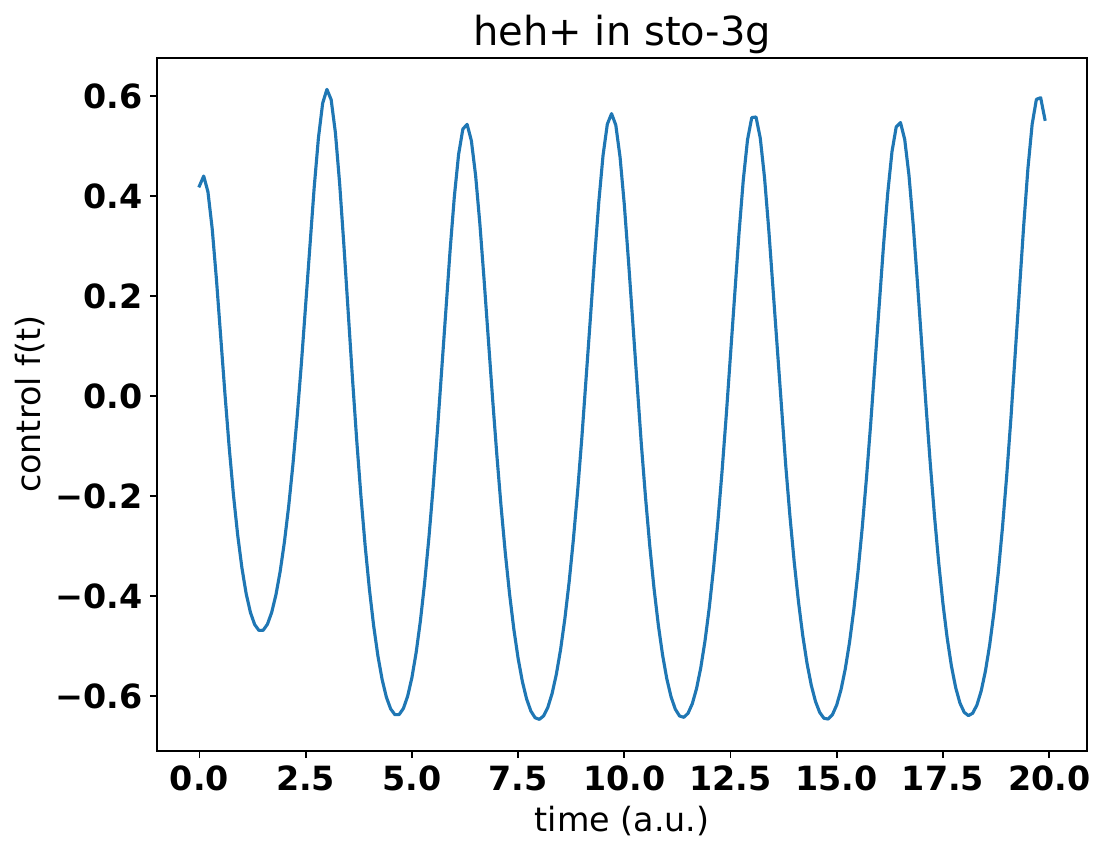} \hspace{-0.2cm} \includegraphics[width=1.795in]{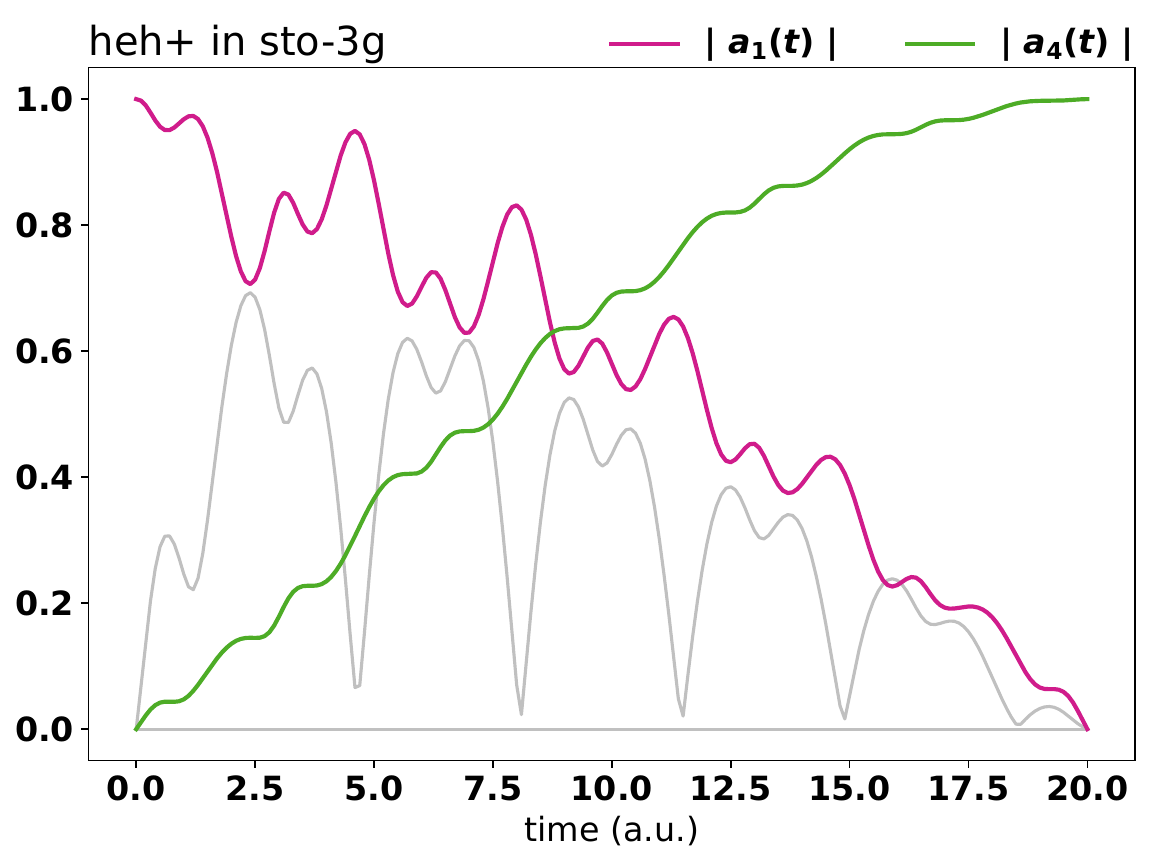} \\
\includegraphics[width=1.705in]{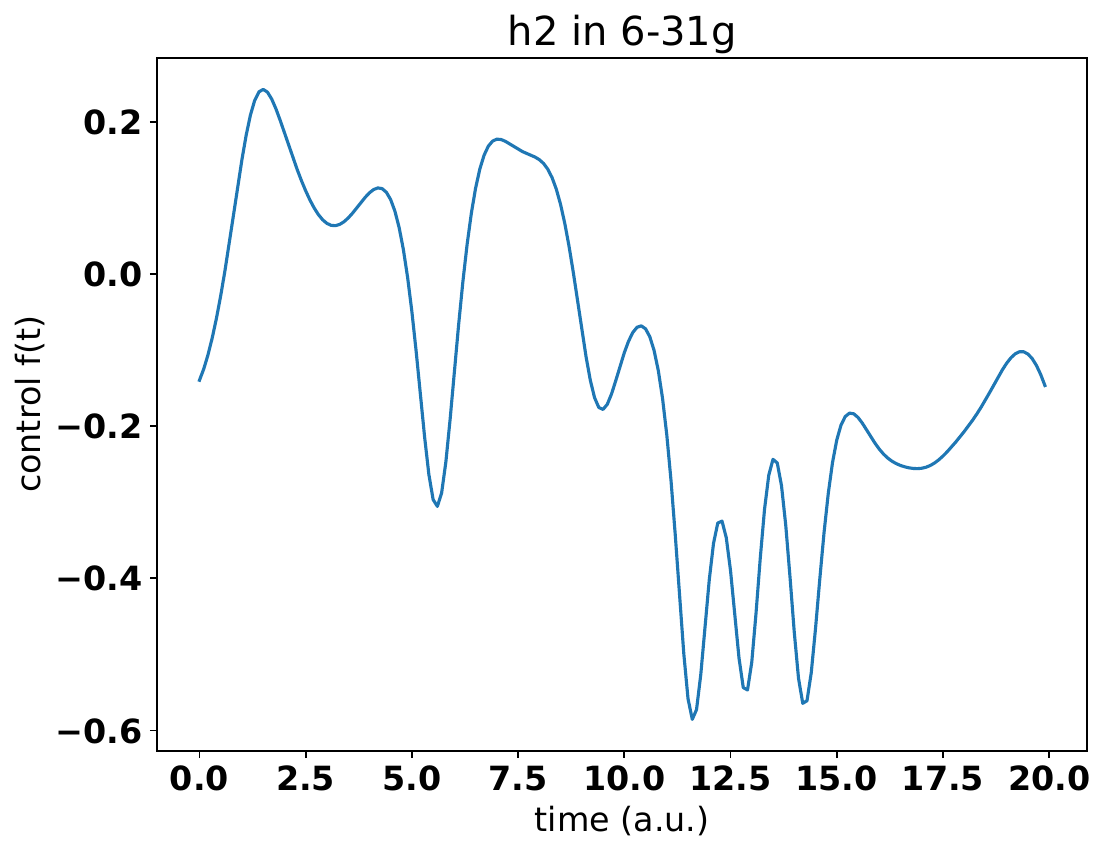}  \hspace{-0.2cm} \includegraphics[width=1.795in]{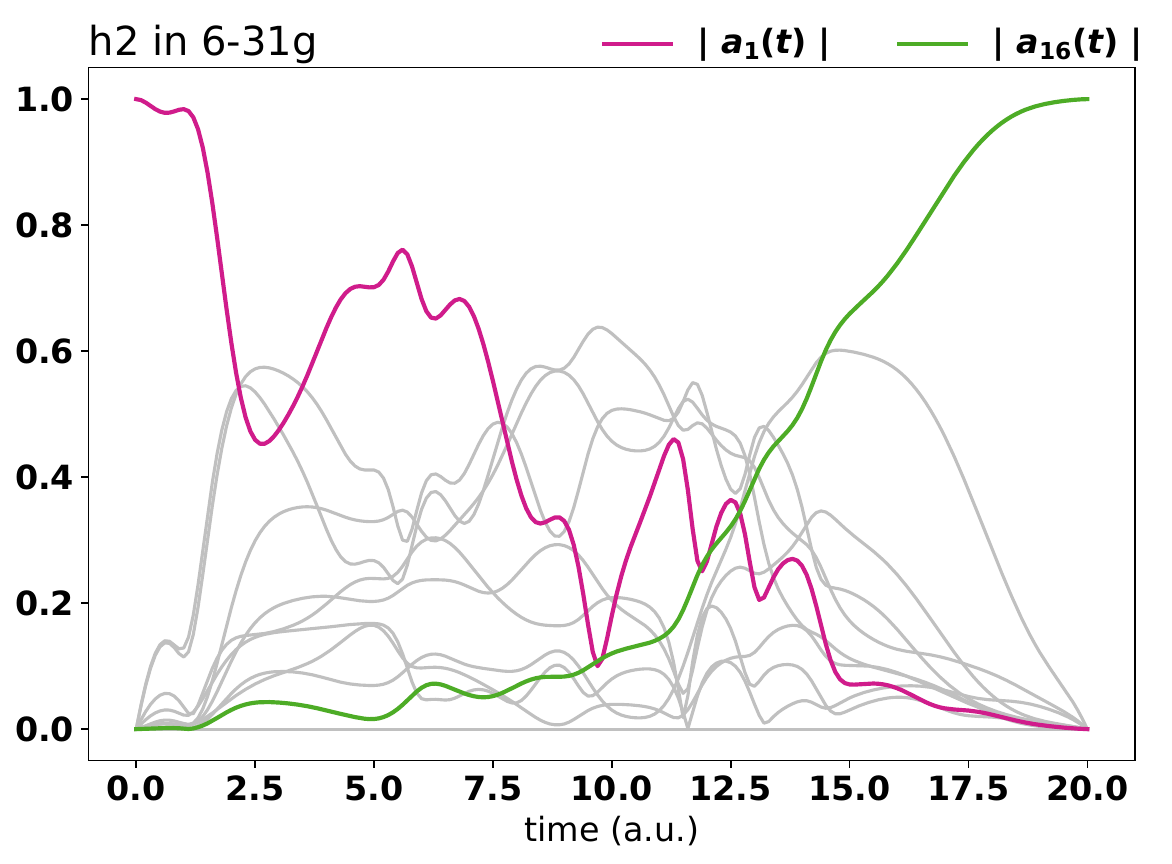} \hspace{-0.3cm} \includegraphics[width=1.705in]{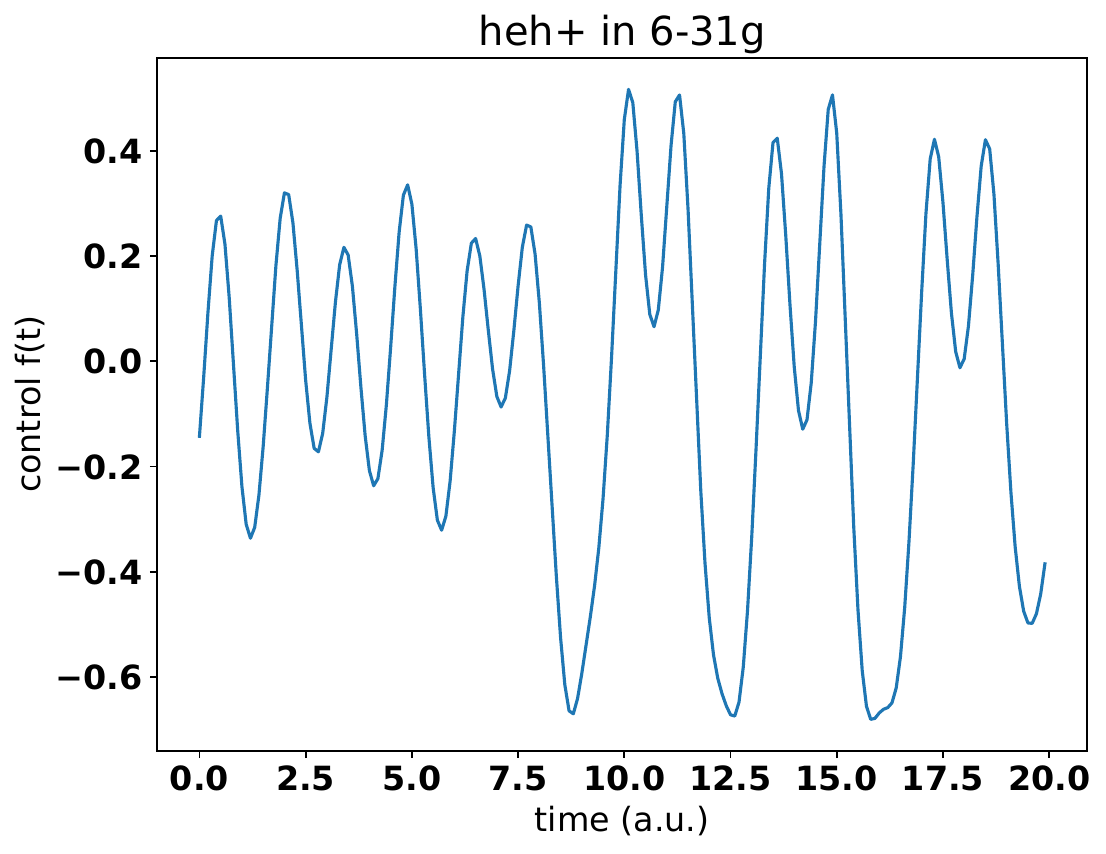} \hspace{-0.2cm}  \includegraphics[width=1.795in]{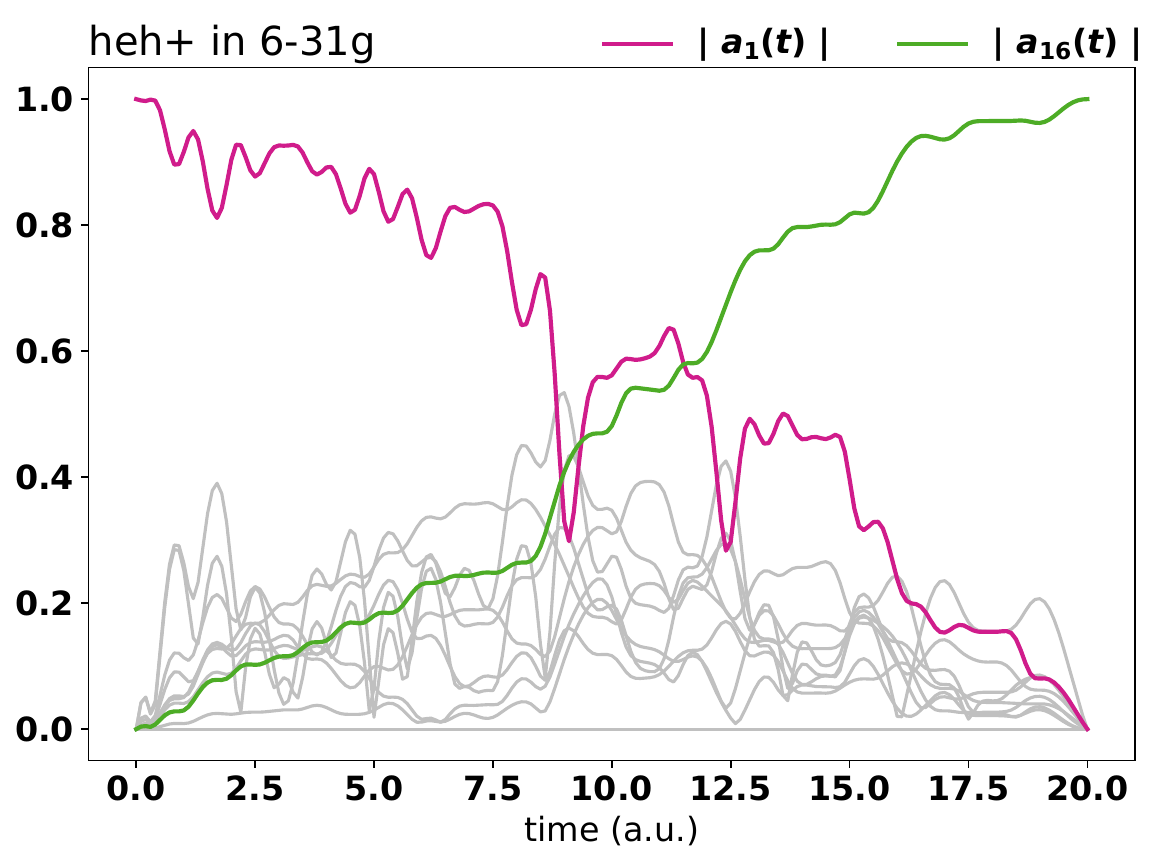}\\}
\caption{From the pool of $1000$ \textbf{maximal model} (\ref{eqn:maximal}) solutions whose statistics were reported in Tables \ref{tab:rawocr}-\ref{tab:ratios}, the plotted solutions achieved the lowest cost.   For each molecular system (choice of molecule plus basis set), we plot both the optimal field strength $f(t)$ and the magnitudes $|a_j(t)|$ of the controlled trajectory.  Here $J = 200$ and $\Delta t = 0.1$.  Gray components of the controlled trajectory have initial and target values of $0$; colored components' initial values differ from their targets.}
\label{fig:bestcompare}
\end{figure*}

\section{RESULTS}
\label{sect:results}
\subsection{Scaling of Gradient and Hessian Algorithms}
\label{sect:scaling}
With the maximal model (\ref{eqn:maximal}), the outer double for loops (over the $\ell$ variable) in Alg. \ref{alg:secondorder} go up to $N_p \geq J$.  Thus a na\"ive implementation of Alg. \ref{alg:secondorder} might be expected to have a run time that scales quadratically in $J$.  On the other hand, vectorization of this outer loop (described in Section \ref{sect:opdiff}) should yield closer-to-linear scaling in $J$.

We test this on systems with four values of $N$, the dimension of the Hamiltonian and dipole moment matrices.  The $N=4$ and $N=16$ systems correspond to full CI for the molecule $\heh$ in, respectively, the STO-3G and 6-31G basis sets.  For this molecular system, only the dipole moment matrix in the $z$ direction is nonzero; hence the control consists of scalars $f_j = f(j \Delta t; \btheta)$, and $\btheta$ is simply a vector of length $J$. For $N=64$ and $N=144$, we created random diagonal core Hamiltonian matrices $H_0$ and random Hermitian dipole moment matrices $M$.   For all systems, we take the initial state to be $\balpha = [1,0,\ldots,0]$ and the target state to be $\bbeta = [0,\ldots,0,1]$.

For a given molecular system, we fix the number of time steps $J$, sample an initial $\btheta^{(0)}$ with each $\theta^{(0)}_j \sim \mathcal{N}(0,1)$ (standard normal), and check the wall clock time to run each algorithm (in turn) on precisely the same set of inputs $\btheta^{(0)}$ and $\{H_0, \{M_k\}_{k=1}^3, \rho, J, \Delta t, \balpha, \bbeta\}$.    We take care to force JIT compilation of our JAX implementations of both algorithms before checking wall clock times.

Fig. \ref{fig:timings} shows a log-log plot of the means and error bars (plus/minus two standard deviations) across $1000$ trials of the above experiment, conducted for each $J = 2^{\kappa}$ with $\kappa = 2, \ldots, 10$.  All tests were conducted on Nvidia A100 GPU nodes on NCSA's Delta GPU cluster; the GPU has 40 GB of RAM. The node also has an AMD Milan CPU with 256 GB RAM \cite{Delta}.  The results  indicate that Alg. \ref{alg:secondorder}'s wall clock time has nearly linear scaling.  For Alg. \ref{alg:firstorder}, the slopes of the OLS (ordinary least squares) regression lines fit to the log-log data for $N=4, 16, 64, 144$ are, respectively, $0.9082$, $0.9067$, $0.9999$, and $1.001$; for Alg. \ref{alg:secondorder}, the slopes are $0.9871$, $0.9938$, $0.9975$, and $1.006$.  Note that $\log_2 y = m \log_2 x + b$ implies $y = 2^b x^m$.  Thus slopes $m \approx 1$ indicate near-linear scaling.

In prior work that did not use second-order adjoints, with a parallel CPU implementation (for $N=144$), we learn that ``Given the same computing resources, a Hessian calculation takes approximately ten times longer than a gradient calculation'' \cite{GoodwinKuprov2016}.  Our results above (esp. for $N=64$ and $N=144$) show that the wall clock time required to obtain the Hessian \emph{and} the gradient (via Alg. \ref{alg:secondorder}) is practically identical to that required to obtain only the gradient (via Alg. \ref{alg:firstorder}).  This shows the advantages of the second-order adjoint method and our vectorized GPU implementation.

\subsection{Optimal Control Results}
\label{sect:ocr}
Continuing with the maximal model (\ref{eqn:maximal}), we consider the question of whether using gradients \emph{and} Hessians improves our ability to solve the optimal control problem, relative to using gradients alone.  We answer this question for both $\heh$ and $\htwo$ in each of two basis sets, STO-3G and 6-31G.  As with $\heh$ (described above), the control consists of an applied electric field in the $z$-direction only; the dipole moment matrices in the $x$- and $y$- directions vanish.  Hence $|\btheta| = J$.  We retain the same initial and target states mentioned above: $\balpha = [1,0,\ldots,0]$ and $\bbeta = [0,\ldots,0,1]$.  For all systems, we use time step $\Delta t = 0.1$, $J=200$ steps, and cost-balance parameter $\rho = 10^6$.  We set the termination criteria tolerances $\delta$ and $\epsilon$ (in Algorithms \ref{alg:firstorder} and \ref{alg:secondorder}) to $10^{-10}$.  We also set the maximum number of iterations to be $10^4$.

Each experiment starts by drawing a random initial guess for $\btheta$, with each entry normally distributed with mean zero and variance one.  With this initial guess, we solve the optimal control problem using Alg. \ref{alg:firstorder}; with the same initial guess, we solve the problem again using Alg. \ref{alg:secondorder}.  We carried out this experiment $1000$ times, recording data on algorithm performance and solution quality.  Optimization always terminated due to satisfaction of either the $\delta$ or $\epsilon$ criteria; the iteration limit was never reached.  In both cases, we couple our algorithms with the trust-region optimizer \emph{trust-constr} in the \emph{scipy.optimize} package \cite{conn2000trust}.  When we use Alg. \ref{alg:firstorder} to supply only gradients to the trust-region optimizer, it uses these gradients to compute approximate inverse Hessians via BFGS updates.  When we use Alg. \ref{alg:secondorder}, the trust-region optimizer uses exact gradients and Hessians.

We measure both algorithms' performance in terms of the iteration count and wall time required to achieve termination criteria.  We measure solution quality in terms of the final cost, the final norm of the gradient (final $\| \text{grad} \|$), and the final norm difference between achieved and desired targets ($\| \text{target viol} \|$).  For each molecule, basis set, and algorithm, we present in Table \ref{tab:rawocr} the mean results across $1000$ trials.  The results show that Alg. \ref{alg:secondorder} requires fewer iterations \emph{and} less wall clock time to achieve solutions whose quality is no worse than those achieved by Alg. \ref{alg:firstorder}.  Based on what we learned from Fig. \ref{fig:timings}, the per-iteration cost of Alg. \ref{alg:secondorder} is nearly identical to that of Alg. \ref{alg:firstorder}, so fewer iterations should imply less wall clock time.  In Table \ref{tab:rawocr}, in each block of two rows (same molecule, same basis set), we highlight in bold the best results.  Whenever Alg. \ref{alg:firstorder}'s result is boldfaced, it is practically equivalent to Alg. \ref{alg:secondorder}'s result.

For each of the $1000$ trials, we also compute the \emph{ratio} of Alg. \ref{alg:firstorder}'s iteration count, wall time, final cost, final $\| \text{grad} \|$, and $\| \text{target viol} \|$ to those of Alg. \ref{alg:secondorder}.  We report in Table \ref{tab:ratios} the means and, in parentheses, the $0.05$ and $0.95$ quantiles of the $1000$ computed ratios in each category.    Let us illustrate how to read Table \ref{tab:ratios} concretely: for $\heh$ in STO-3G, Table \ref{tab:ratios} reports that Alg. \ref{alg:firstorder} will require (on average across $1000$ trials) $5.08$ times as many iterations than Alg. \ref{alg:secondorder} to achieve the termination criteria.  For this molecular system, there is an empirical probability of $0.90$ that Alg. \ref{alg:firstorder} required between $2.12$ and $9.88$ times more iterations than Alg. \ref{alg:secondorder} to achieve the termination criteria.

The means and quantile intervals in Table \ref{tab:ratios} confirm that for most choices of the initialization $\btheta^{(0)}$, Alg. \ref{alg:secondorder} converges more rapidly to solutions of substantially the same quality as Alg. \ref{alg:firstorder}.  While there do exist bad initializations $\btheta^{(0)}$ that lead to Alg. \ref{alg:secondorder} producing a worse final solution than Alg. \ref{alg:firstorder} (in terms of cost and/or $\| \text{target viol} \|$), this is uncommon.  At least 90\% of the time, both algorithms' final values of cost and $\| \text{target viol} \|$ are  within an order of magnitude.

For each of our four molecular systems, we use the final value of the optimized cost to select the best solution $\btheta$ obtained across the $1000$ trials.  In Fig. \ref{fig:bestcompare}, we plot these optimal control signals $f(t) = f_3(t)$ (left panels) and the magnitudes $|a_j(t)|$ of the corresponding controlled trajectories (right panels). As we have used the maximal model (\ref{eqn:maximal}), what we actually plot here (in the left panels) is $f(j \Delta t; \btheta) = \theta_j$ as a function of $j$.  Without any regularization to enforce smoothness in time of the control $f(t)$, the resulting solutions are all smooth, with a visual resemblance to sums of sinusoids.  In all cases, the controlled trajectories achieve their targets with an error on the order of $10^{-5}$.

Consider again the best maximal model solutions produced by Alg. \ref{alg:secondorder} and plotted in Fig. \ref{fig:bestcompare}; for each molecular system, the plotted $f(t)$ and $|a_j(t)|$ curves correspond to a particular $\btheta$ vector.  When we examine the eigenvalues of the Hessians corresponding to these four theta vectors (one per molecular system), we find that all eigenvalues are positive.  Together with the near-vanishing gradients for Alg. \ref{alg:secondorder} reported in Table \ref{tab:rawocr}, we conclude that with the maximal model (\ref{eqn:maximal}), our methods have succeeded in finding locally optimal controls.

\section{CONCLUSION}
Our results show the utility of the second-order adjoint method and Alg. \ref{alg:secondorder} in terms of (i) superior algorithm performance (compared with a first-order method) and (ii) allowing for arbitrary parameterizations of the control $\bbf(t)$.  In future work, we aim to impose further constraints on $\bbf(t)$ to ensure its experimental realizability, as in \cite{Larocca2020}.  We also aim to explore whether checkpointing ideas (e.g., as in \cite{Narayanan2022}) can be used to reduce the memory requirements of Alg. \ref{alg:secondorder}.

\end{document}